\documentclass[pra,aps,twocolumn,showpacs,groupauthors,nofootinbib]{revtex4}

\usepackage{amssymb}
\usepackage[english]{babel}
\usepackage[ansinew]{inputenc}
\usepackage{amsfonts}
\usepackage{amsmath}
\usepackage{graphicx}
\usepackage{bm}
\usepackage{color}

\newcommand{\ket}[1]{|{#1}\rangle}
\newcommand{\bra}[1]{\langle{#1}|}

\begin{document}


\title{Dynamical programming of continuously observed quantum systems}


\author{Viacheslav~P.~Belavkin$^1$}
\author{Antonio~Negretti$^2$}
\email[E-mail: ]{antonio.negretti@uni-ulm.de}
\altaffiliation[Present address: ]{Institute for
Quantum Information Processing, University of Ulm, Albert-Einstein-Allee 11, D-89069 Ulm,
Germany.}
\author{Klaus~M{\o}lmer$^2$}
\affiliation{1. School of Mathematics, Nottingham University, Nottingham, NG7 2RD, United
Kingdom\\
2. Lundbeck Foundation Theoretical Center for Quantum System Research\\
Department of Physics and Astronomy, Aarhus University\\
DK-8000 Aarhus C, Denmark}
\date{\today}


\begin{abstract}
We develop dynamical programming methods for the purpose of optimal
control of quantum states with convex constraints and concave cost
and bequest functions of the quantum state. We consider both open
loop and feedback control schemes, which correspond respectively to
deterministic and stochastic Master Equation dynamics. For the
quantum feedback control scheme with continuous non-demolition
observations we exploit the separation theorem of filtering and
control aspects for quantum stochastic dynamics to
derive a generalized Hamilton-Jacobi-Bellman equation. If the
control is restricted to only Hamiltonian terms this is equivalent
to a Hamilton-Jacobi equation with an extra linear dissipative term.
In this work, we consider, in particular, the case when control is
restricted to only observation. A controlled qubit is considered as
an example throughout the development of the formalism. Finally, we
discuss optimum observation strategies to obtain a pure state from a
mixed state of a quantum two-level system.
\end{abstract}




\maketitle


\section{Introduction}

The dynamical theory of quantum nondemolition observation, developed
by Belavkin in the 80's
\cite{Belavkin1980,Belavkin1985,Belavkin1987} resulted in a new
class of quantum stochastic equations
\cite{Belavkin1988,Belavkin1989a,Belavkin1989b,Belavkin1990} for
quantum dissipative systems under observation. Different quantum
jump and quantum diffusive stochastic equations are obtained when
the observed quantity has discrete and continuous spectra as, e.g.,
in photon counting and homodyne detection of optical fields. In the
last case, for linear models and initial quantum Gaussian states
this allows explicit solutions in terms of a quantum analog of the
Kalman linear filter first introduced in Ref.\cite{Belavkin1980}
(see also \cite{Belavkin1987}, \cite{Belavkin1999},
\cite{Belavkin1989c,Belavkin1992a,Belavkin1993b,Belavkin1994b}).
These equations are similar to the classical filtering equation
derived by Stratonovich in the 60's \cite{Stratonovich1960} for
classical partially observed conditional Markov systems.

In the classical theory of feedback control the so called
\textit{Separation Theorem}
\cite{Stratonovich1962,Stratonovich1968,Davis1977} applies. This
theorem states that the full control problem can be reduced in two
separated and independent parts: $(i)$ the estimation of the state
of the system; $(ii)$ the optimal control of the system. The same
approach can be applied to quantum systems, as first pointed out by
Belavkin in Ref.\cite{Belavkin1983} and more recently implemented in
the quantum dynamical programming method
\cite{Belavkin1999,Belavkin2005,Gough2005}. The only difference
between classical and quantum systems is related to the concept of
``state'' given in the two theories. In both mechanics the state
carries the necessary information to describe fully the system and
this characterizes our knowledge of it. In classical mechanics a
system is usually described by its position and momentum 
phase space variables. In quantum mechanics a state is described by
a state vector which belongs to a linear Hilbert space or by a von
Neumann density matrix. Contrary to the classical formulation, the
linear property of the Hilbert space allows superpositions of
quantum states. This, however, does not play a role in the
separability of the state estimation and optimal control problems.
The optimal control of the state of a quantum system implies the
control of a density matrix valued stochastic process and is
mathematically equivalent to any classical control problem.

Experimentally, very important achievements have been obtained in 
the last decade which have led to exciting prospectives to manipulate 
quantum systems. For instance, high precision quantum measurements at 
the Heisenberg limit have been implemented \cite{Armen2002,Geremia2003,Cook2007}
and quantum feedback has been used to record an externally provided 
quantum state of light onto an atomic ensemble \cite{Julsgaard2004}. 
At the same time several theoretical proposals for quantum state
engineering like spin-squeezing
\cite{Molmer1999,Thomsen2002,Madsen2004,Nielsen2008}, 
photon number states \cite{Geremia2006}, entangled states 
\cite{Sherson2005,Yanagisawa2006a,Mancini2006}, quantum superposition states
of optical fields \cite{Negretti2007} and of two macroscopically
distinguishable atomic states \cite{Nielsen2009}, and cooling 
of either a mechanical resonator \cite{Mancini1998}
or the atomic motion in an optical cavity \cite{Steck2004} via
continuous measurement have been put forward. Among these activities 
more related mathematical issues as stability and observability become 
relevant and have been subject of investigation \cite{Belavkin2005,Gough2005}, 
\cite{vanHandel2005,Doherty2000,James2004,Yanagisawa2006,Jacobs2008}.

We shall follow the approach of Ref.\cite{Gough2005} (which is also taken 
in most papers on quantum feedback control after Wiseman \cite{Wiseman1994}), 
where the filtering controlled equation is postulated but not derived as a 
result of the conditioning of quantum dynamics upon the nondemolition
observations. To simplify the mathematics, we shall consider only
finite-dimensional models (like qubits). However, unlike in \cite{Gough2005}, 
we shall consider optimization of not only Hamiltonian
control but also of control exercised by the choice among different 
observations carried out on the system. In the original setup 
\cite{Belavkin1987,Belavkin1988}, \cite{Belavkin1983}, of optimal
quantum feedback control theory the cost and target functions were 
restricted to affine functionals of the state where these functions 
are therefore expectations of certain observables. The result
of minimization of an affine function as expected cost of a controlled 
observable is not always affine but always concave. This point was 
argued in Ref.\cite{Fuchs2001}, and the use of nonlinear concave
cost functions was justified by Jacobs \cite{Jacobs2003}, Wiseman and 
Ralph \cite{Wiseman2006}, Wiseman and Bouten \cite{Wiseman2008} for 
optimal feedback control of qubit purification. We will thus consider
concave cost and target functions of the properties of the system controlled 
by the feedback.

The paper is organized as follows. In Sec. \ref{sec:sgd}, we
introduce the density matrix notation, time evolution generators,
and the notion of derivatives with respect to a quantum state. In
Sec. \ref{sec:HJB-det}, we introduce Bellman and Hamilton-Pontryagin
optimality, and we introduce cost and bequest functions. In Sec.
\ref{sec:HJB-obs}, we turn to the problem of quantum dynamics under
observation and present the diffusive quantum filtering equation
corresponding to homodyne or heterodyne measurements. In Sec.
\ref{sec:ofc}, we analyze the optimal control problem with
observations and constraints, and we derive a Bellman equation for
filtered dynamics. In Sec. \ref{sec:oep}, we discuss the special
case of purification of a mixed quantum state by measurements. We
conclude the paper with a discussion in Sec. \ref{sec:disc}.


\section{States, generators and derivatives}
\label{sec:sgd}

We will assume a complex, finite dimensional, Hilbert space $\mathfrak{h} =
\mathbb{C}^d$ for our open (observed and controlled) quantum system. It is 
convenient to define the quantum state space $\mathcal{S}$ as the compact, convex set 
of positive and hermitian density matrices, normalized with respect to the 
unit trace, $\mathrm{tr}\{\varrho\} \equiv \frac{1}{d} \mathrm{Tr}\{\varrho\} = 1$.

The real linear combinations of such matrices form the linear space
$\mathcal{L}$ of Hermitian matrices. In general every state $\varrho
$ can be parametrized as $\varrho \left( q\right) =\varrho _{0}-q$
with respect to a given state $\varrho _{0}\in \mathcal{S}$ by a
corresponding $q\in \mathcal{L}_{0}$, where $\mathcal{L}_{0}\subset
\mathcal{L}$ is the space of trace zero matrices.

\textbf{Example}: Throughout this text, we shall illustrate and apply our
results and formalism to the example of a single two-level quantum system,
also described as a qubit. Any qubit density matrix with respect to the
normalized trace $\mathrm{tr}\{\cdot \}:=\frac{1}{2}\mathrm{Tr}\{\cdot \}$
can be expanded on the Identity and the three Pauli matrices,
$\varrho =I+\sigma _{\vec{r}}$, with
$\sigma _{\vec{r}}=r_{x}\sigma _{x}+r_{y}\sigma _{y}+r_{z}\sigma _{z}\equiv
\vec{r}\cdot \vec{\sigma}$\footnote{For vectors in $\mathbb{R}^{3}$ we use
the arrow notation, e.g., $\vec{r}$. To avoid misunderstandings and to simplify
the notation we use bold Latin characters to indicate the projection $\mathbf r$
of a three dimensional vector $\vec{r}$ on a subspace
$\mathbb{R}^{d}\subseteq \mathbb{R}^3$ with $d\leq 3$.}.
The quantum state space $\mathcal{S}$ thus becomes associated with the
unit ball (Bloch sphere) $\left\{\vec{r}:\left\vert \vec{r}\right\vert \leq 1\right\}$.

For the qubit we have $q=-\sigma _{\vec{r}}$ with respect to the
central $\varrho _{0}=I$. Thus $q\left(\varrho \right) $ can be identified with the
Euclidean vector $\vec{q}=-\vec{r}$ globally parametrizing $\varrho \in \mathcal{S}$
as $\varrho =I-\sigma _{\vec{q}}$.

The quantum state Master Equation (ME) describing an open quantum system is defined as

\begin{equation}
\frac{\mathrm{d}}{\mathrm{d}t}\varrho ^{t} + \upsilon \left( u(t),\varrho^t \right) = 0,
\label{master equation}
\end{equation}
where $\varrho^t\in\mathcal{S}$ and the generator or drift term $\upsilon$ is given by

\begin{align}
\upsilon\left( u,\varrho\right) & =\frac{\mathrm{i}}{\hbar }\left[ H\left(
u\right) ,\varrho\right] +\sum_{j}\upsilon_{L_j(u)}\left(\varrho\right),
\label{generator} \\
\upsilon_{L_j}\left(\varrho\right) & =
\frac{L^{j\dagger }L^{j}\varrho + \varrho L^{j\dagger }L^{j}}{2}
- L^{j}\varrho L^{j\dagger },
\label{lindblad}
\end{align}
with $H\left( u\right) =$ $H\left( u\right) ^{\dagger }$ self adjoint, and
$L^{j}\left( u\right)$ belongs to the  complex space 
$\mathcal{B}\left( \mathfrak{h}\right)$ of bounded operators on 
$\mathfrak{h}$ for each value of the set of control parameters $u$. The parameter
$u=u\left( t\right)$ is the admissible control trajectory which may be restricted to a domain
$\mathcal{U}(t) \subseteq\mathbb{R}^{n}$ of dimensionality $n$, possibly depending on time $t$.

\textbf{Example}: Assume that the qubit dissipative dynamics is given by a
Hamiltonian $H\left( \mathbf{u}\right) =\frac{\hbar }{2}\sigma _{\mathbf{u}}$
controlled by the magnetic field $\mathbf{u}\left( t\right) \in \mathbb{R}%
^{d}$ for $d\leq 3$, and by a single dissipation operator $L=\frac{1}{2}%
\lambda \sigma _{z}$. With the Pauli matrix representation $\varrho
=I-\sigma _{\vec{q}}$ for the density matrix, the ME evolution is
governed by
\begin{equation}
\upsilon\left( \varrho , \mathbf{u}\right) =\vec{\sigma}\cdot \left(
\mathbf{u}\times \vec{q}\right) - \frac{|\lambda |^{2}}{2}\left( q_{x}\sigma
_{x}+q_{y}\sigma _{y}\right).
\label{qubitlingen}
\end{equation}

A (nonlinear) functional $\varrho\mapsto\mathsf{F}\left[ \varrho\right] $
admits a derivative if there exists a $\mathcal{B}\left( \mathfrak{h}\right)$-valued
function $\nabla_{\varrho }\mathsf{F}\left[ \cdot\right] $ such that
\begin{equation}
\lim_{h\rightarrow0}\frac{1}{h}\left\{ \mathsf{F}\left[ \cdot+h\tau\right] -
\mathsf{F}\left[ \cdot\right] \right\} =\left\langle \tau,\nabla_{\varrho }
\mathsf{F}\left[ \cdot\right] \right\rangle \;\;\;\;\forall\tau\in\mathcal{L}_{0},
\label{nabla_Q}
\end{equation}
where we have introduced the pairing
$\left\langle \varrho,X\right\rangle :=\mathrm{tr}\left\{ \varrho X\right\}$
with $\varrho\in\mathcal{S}$ and $X\in\mathcal{L}^{\star}$, where
$\mathcal{L}^{\star} = \mathcal{B}\left( \mathfrak{h}\right)$ is the adjoint 
space of $\mathcal{L}$.

If $\varrho^{t}$ is a quantum state trajectory controlled by equation
(\ref{master equation}), then we may apply the chain rule

\begin{equation*}
\frac{\mathrm{d}}{\mathrm{d}t}\mathsf{F}\left[ \varrho^{t}\right]
=- \left\langle \upsilon\left( u\left( t\right) ,\varrho
^{t}\right) ,\nabla_{\varrho}\mathsf{F}\left[ \varrho^{t}\right]
\right\rangle
\end{equation*}
for such a functional $\mathsf{F}$.

A Hessian $\nabla_{\varrho}^{\otimes2}\equiv\nabla_{\varrho}\otimes\nabla_{\varrho}$
is defined as

\begin{equation*}
\lim_{h\rightarrow0}\frac{1}{h}\left\langle \tau^{\prime},\nabla_{\varrho }%
\mathsf{F}\left[ \varrho+h\tau\right] -\nabla_{\varrho }%
\mathsf{F}\left[ \varrho\right] \right\rangle=\left\langle \tau\otimes
\tau^{\prime},\nabla_{\varrho}^{\otimes2}\mathsf{F}\left[ \varrho\right]
\right\rangle ,
\end{equation*}
with $\tau,\tau^{\prime}\in\mathcal{L}_0$ and we say that the functional is twice
continuously differentiable whenever $\nabla_{\varrho}^{\otimes2}\mathsf{F}\left[ \varrho\right]$
exists.

\textbf{Example}: Let $\mathsf{F}\left[ \varrho \right] =f\left[ \vec{q}\right] $ be a
smooth function of the state, i.e., of $\vec{q}$. Then $\nabla _{\varrho }%
\mathsf{F}\left[ \varrho \right] $ can be directly identified with $-\vec{%
\sigma}\cdot \vec{\nabla}f\left( \vec{q}\right) $ in the sense that $%
\left\langle \tau ,\nabla _{\varrho }\mathsf{F}\left[ \varrho \right]
\right\rangle =-\vec{t}\cdot \vec{\nabla}f\left( \vec{q}\right) $ for any $%
\tau =\sigma _{\vec{t}}\in \mathcal{L}_{0}$.
Here the minus sign is related to the fact that the state $\varrho$ is identified
with $\vec{r}$, but the gradient $\vec{\nabla}f\left( \vec{q}\right)$ is considered
with respect to $\vec{q}=-\vec{r}$.

Similarly we can write
\begin{equation*}
\nabla _{\varrho }^{\otimes 2}\mathsf{F}\left[ 1-\sigma _{\vec{q}}\right]
=\left( \vec{\sigma}\cdot \vec{\nabla}\right) ^{\otimes 2}f\left( \vec{q}\right).
\end{equation*}


\section{Bellman and Hamilton-Pontryagin Optimality}
\label{sec:HJB-det}

\subsection{Cost functions}

Let us consider the integral cost for a control function $\left\{
u\left( t\right) \right\} $ of the quantum state $\varrho ^{t}$ over
a time-interval $(t_{0},T]$
\begin{equation}
\mathsf{J}\left[ \left\{ u\left( t\right) \right\} ;t_{0},\varrho _{0}\right]
=\int_{t_{0}}^{T}\mathsf{C}\left( u\left( t\right) ,\varrho ^{t}\right)
\mathrm{d}t+\mathsf{G}\left( u\left( T\right) ,\varrho ^{T}\right) ,
\label{J}
\end{equation}
where $\left\{ \varrho ^{t}:t\in (t_{0},T]\right\} $ is the solution
to a quantum controlled ME with initial condition
$\varrho^{t_{0}}=\varrho _{0}$, and $\mathsf{C}$ is a \textit{cost
density}  while $\mathsf{G}$ is the \textit{terminal cost, or
bequest function.} 
Causality implies that for any $t \in [0,T]$ the state $\varrho^t$ depends only
on $u(t^{\prime})$ with $t^{\prime}\in [0,t)$ and is independent of the current and future
values of $u(t)$. In particular, the choice of
$u(T)$ at the terminal time instant $T$ does not affect the state $\varrho^T$. We
emphasize that the admissible control strategies $u(t)$ are not necessarily
continuous but they can be assumed right continuous for all $t$ with left
limits $u(t_-)$ not necessarily equal to $u(t)$. One can thus, for any $t$, regard
$u(t)$ as entirely separate from earlier values $u(t^{\prime}<t)$, while any later value
$u(t^{\prime\prime} \geq t)$, serves as a "postprocessing" control for $\varrho^t$. Our task
is to adapt $u(t)$ to minimize the contribution from the cost density and to
use $u(T)$ to "postprocess" the final quantum state or to modify the terminal
cost or bequest function in Eq.(\ref{J}) to most successfully achieve the desired
goal. This will be exemplified in the following.

\subsection{Quantum dynamical programming}

Let us first consider the quantum optimal control theory without
observation, assuming that the state $\varrho^{t}\in\mathcal{S}$
obeys the ME (\ref{master equation}). To identify the optimal
control strategy $\left\{ u\left( t\right) \right\} $ with the
specific cost $\mathsf{J}\left[ \left\{ u\right\}
;t_{0},\varrho_{0}\right] $, we note that for times $t<t+h<T$, one
has
\begin{align*}
\mathsf{S}\left( t,\varrho\right) &:=\inf_{\left\{u\right\}} \,\left\{
\int_{t}^{t+h}\mathsf{C}\left( u\left( r\right) ,\varrho^{r}\right) \mathrm{d%
}r\right. \\
&\phantom{=} \left.+\int_{t+h}^{T}\mathsf{C}\left( u\left( r\right)
,\varrho^{r}\right) \mathrm{d}r+\mathsf{G}\left( u\left( T\right)
,\varrho^{T}\right) \right\}.
\end{align*}

Now, we assume that $\left\{ u^{\mathrm{o}}\left( r\right) :r\in(t,T]\right\}$ 
is the optimal control when starting in state $\varrho$ at time $t$, and
denote by $\left\{ \varrho^{r}:r\in(t,T]\right\}$
the corresponding state trajectory $\varrho^{r}=\varrho^{r}\left( t,\varrho\right)$.
According to Bellman's optimality principle\footnote{The optimality principle
  states that if the path of a process through the stages $t_a\rightarrow
  t_b\rightarrow t_c$ is the optimal path from $t_a$ to $t_c$, then the path 
  from $t_b$ to $t_c$ is optimal as well
  \cite{Kirk2004,Davis1977,Bellman1957}.} 
the control $\left\{u^{\mathrm{o}}\left( r\right) :r\in(t+h,T]\right\} $ is
then optimal for the evolution starting from $\varrho^{t+h}$ at the later 
time $t+h$, and hence
\begin{equation*}
\mathsf{S}\left( t,\varrho\right) =\inf_{\left\{ u\right\}
}\,\left\{ \int_{t}^{t+h}\mathsf{C}\left( u\left( r\right)
,\varrho^{r}\right) \mathrm{d}r+\mathsf{S}\left( t+h,\varrho^{t+h}\right)
\right\} .
\end{equation*}
For $h$ small we expand $\varrho^{t+h}=\varrho-\upsilon\left(
u\left( t\right) ,\varrho\right) h+o\left( h\right) $ and we may
apply a Taylor expansion of $\mathsf{S}\left( t,\varrho\right)$.
Then, by taking the limit $h\rightarrow0$ we obtain \cite{Gough2005}
\begin{equation}
-\frac{\partial}{\partial t}\mathsf{S}\left( t,\varrho\right) =\inf
_{u}\left\{ \mathsf{C}\left( u,\varrho\right) -\left\langle \upsilon\left(
u,\varrho\right) ,\nabla_{\varrho}\mathsf{S}\left( t,\varrho\right)
\right\rangle \right\}.
\label{HJB1}
\end{equation}
This equation should be solved subject to the terminal condition
\begin{equation}
\mathsf{S}\left( T,\varrho\right) =\inf_{u}\left\{ \mathsf{G}\left(
u,\varrho\right) :u\in\mathcal{U}\left( T\right) \right\} \equiv \mathsf{S}%
_{T}[\varrho].
\end{equation}
We recall that when the infimum is reached, then the objective of
the optimization problem we aim to solve is obtained. This formalism
has been applied in the case of optimal control of the cooling of a
quantum dissipative three-level $\Lambda$ system \cite{Sklarz2004},
and in the classical thermodynamic optimization of the evolutionary
Carnot problem \cite{Sieniutycz1997}.

\subsection{Quantum Pontryagin Hamiltonian}

We introduce the Pontryagin Hamiltonian function defined, for $q\in\mathcal{L}_0$,
$p\in\mathcal{L}^{\star}$, by

\begin{equation}
\mathsf{H}_{\upsilon }\left( q,p\right) :=\sup_{u}\left\{ \left\langle
\upsilon \left( u,\varrho \left( q\right) \right) ,p\right\rangle -\mathsf{C}
\left( u,\varrho \left( q\right) \right) \right\}.
\label{hamiltonian}
\end{equation}
We use the parametrization $\varrho \left( q\right) =\varrho _{0}-q$
by a zero trace operator $q\in \mathcal{L}_{0}$ and $\upsilon $ is
the velocity $\dot{q}$ of $q=\varrho _{0}-\varrho $. The
equations of motion for the state operator $q$ and for the operator
$p$ can be expressed with the Pontryagin Hamiltonian in formally the
same way as the equations of motion for the canonical coordinates
$(q,p)$ in the Hamiltonian formulation of classical mechanics
\cite{Gough2005}.

Since $\left\langle \upsilon \left( u,\varrho \right)
,I\right\rangle =0$, the Pontryagin Hamiltonian does not change if
we replace  any $p\in \mathcal{L}^{\star }$ by $p+\lambda I$ with
$\lambda \in \mathbb{C}$. The mathematical consequences of this
equivalence class property and the observation that the operator
 $p$ in Eq.(\ref{hamiltonian}) is the Legendre-Fenchel transform of the
 cost function $\mathsf{C}$ are further developed in Ref.\cite{Gough2005}.

We may use the Pontryagin Hamiltonian to rewrite (\ref{HJB1}) as the (backward)
\textit{Hamilton-Jacobi-Bellman} (HJB) equation
\begin{equation}
-\frac{\partial }{\partial t}\mathsf{S}\left( t,\varrho \left( q\right)
\right) +\mathsf{H}_{\upsilon }\left( q,p\left( \nabla_{\varrho } \mathsf{S}\left(
t,\varrho \right) \right) \right) =0,
\label{HJB2}
\end{equation}
which can be simply written as $\partial _{t}\mathsf{S}\left( t,\varrho
\right) =\mathsf{H}_{\upsilon }\left( \varrho _{0}-\varrho ,p\left( \nabla_{\varrho } \mathsf{S}%
\left( t,\varrho \right) \right) \right) $ for any a priori chosen reference state $%
\varrho _{0}\in \mathcal{S}$.

\textbf{Example:} In the case of the Hamiltonian controlled dissipative
dynamics (\ref{qubitlingen}) we have with $p = \sigma_{\vec{p}} + \mathbb{C}I$
\begin{equation*}
\left\langle \upsilon ,p\right\rangle =\left( \vec{q}\times \vec{p}\right)
\cdot \mathbf{u-}\frac{|\lambda |^{2}}{2}\left( q_{x}p_{x}+q_{y}p_{y}\right).
\end{equation*}
For the qubit with the density cost

\begin{equation*}
\mathsf{C}\left( \mathbf{u},\varrho\right) =O_{B_{1}}^{+}\left( \mathbf{u}\right),
\;\;O_{B_1}^{+}\left( \vec{u}\right) = \left\{
\begin{array}{l}
0,\;\;\;\;\vec{u}=\mathbf{u}\in B_1 \vspace{0.2cm} \\
+\infty ,\;\;\;\;\vec{u}\notin B_1
\end{array}
\right.
\end{equation*}
under the constraint $B_{1}=\left\{ \mathbf{u}\in \mathcal{U}:\left\vert \mathbf{u}\right\vert
\leq 1\right\}$, the supremum in Eq.(\ref{hamiltonian}),

\begin{align*}
\mathsf{H}_{\upsilon }\left( q,p\right) &:= \sup_{|\mathbf{u}|\leq
1}\left\{ \left\langle \upsilon \left( u,\varrho \left( q\right) \right)
,p\right\rangle \right\} \\
\phantom{=} &=\sup_{|\mathbf{u}|\leq 1}\{\mathbf{u\cdot }\left( \vec{q}\mathbf{\times }
\vec{p}\right) \}-\frac{|\lambda |^{2}}{2}\left( q_{x}p_{x}+q_{y}p_{y}\right),
\end{align*}
is achieved at the stationary point $\mathbf{u}^{\mathrm{o}}\left( \vec{q}\right)=
\mathbf{p}\left( t,\vec{q}\right) /\left\vert \mathbf{p}\left( t,\vec{q}\right) \right\vert $,
with $\mathbf{p}=\left( \vec{q}\times \vec{p}\right) _{\mathcal{U}}$ denoting the projection
of $\vec{q}\times \vec{p}$ onto $\mathcal{U}$, where the costate $\vec{p}\left( t,\vec{q}
\right)=-\nabla _{\vec{q}}s\left( t,\vec{q}\right)$ is obtained from the solution
$s\left(t,\vec{q}\right) =\mathsf{S}[\varrho \left( \vec{q}\right) ]$ of the HJB equation (\ref{HJB2}).
This yields the Pontryagin Hamiltonian with dissipation

\begin{equation*}
\mathsf{H}_{\upsilon }\left( q,p\right) =\left\vert \mathbf{p}\left( t,\vec{q
}\right) \right\vert +\frac{|\lambda |^{2}}{2}\left( q_{x}\frac{\partial }{
\partial q_{x}}+q_{y}\frac{\partial }{\partial q_{y}}\right) s\left(t,\vec{q}
\right).
\end{equation*}

\subsection{Linear, affine, and concave cost and bequest functions}

Now, we consider cost density and  bequest functions which are
linear functions of the quantum state $\varrho$

\begin{equation}
\mathsf{C}\left( u,\varrho \right) =\left\langle \varrho ,C\left( u\right)
\right\rangle ,\;\mathsf{G}\left( u,\varrho \right) =\left\langle \varrho
,G\left( u\right) \right\rangle,
\label{costs}
\end{equation}
\textit{i.e.}, they can be interpreted as the expectation values of a \textit{cost observable}
$C\left( u\right)$  and a \textit{bequest observable} $G\left( u\right)$, which may depend on
the control parameter $u\in \mathbb{R}^{n}$. One can consider, for example, an average energy
associated with the control parameter as a cost, say $\mathsf{C}\left(u\right) =u^{2}/2$, and
the error probability $\mathsf{G}=\left\langle\varrho ,I-P_{T}\right\rangle $ given by the
orthoprojector $P_{T}=\ket{\psi_{T}}\bra{\psi _{T}}$ on a target state-vector $\ket{\psi _{T}}$
as the bequest observable. Although the dependence of $\mathsf{G}$ on the final value $u\left( T\right)$
of the control parameter in Eq.(\ref{J}) is sometimes redundant, for the sake of generality
and for reasons which will be clear below we keep this dependence.

It is natural to extend the cost and bequest functions  to \textit{affine}
functions
\begin{eqnarray}
\mathsf{C}\left( u,\varrho \right) &=&\left\langle \varrho ,C\left( u\right)
\right\rangle +c\left( u\right),  \notag \\
\mathsf{G}\left( u,\varrho \right) &=& \left\langle \varrho ,G\left(
u\right) \right\rangle +g\left( u\right),  \label{affcosts}
\end{eqnarray}
of the state $\varrho $ which can be obtained from the linear functions by
replacing $C\mapsto C+cI$ and $G\mapsto G+gI$ in (\ref{costs}).
This generalization has no consequences, unless the real-valued functions $c\left( u\right)$
and $g\left( u\right)$ are allowed to take also the infinite cost value
\thinspace $+\infty $, thus reflecting rigid \textit{constraints} on $u$. Indeed,
any constraint on the admissible domain $\mathcal{U}\left( t\right)$,
can be described by the infinite costs
\begin{equation*}
c\left( t,u\right) =\infty \;\;\forall u\notin \mathcal{U}\left( t\right)
,\;\;\;g\left( u\right) =\infty \;\;\forall u\notin \mathcal{U}\left(
T\right) ,
\end{equation*}
such that the expected cost is finite only if $u\left( t\right) $ is
in the allowed domain $\mathcal{U}\left( t\right)$. For instance, in
the deterministic preparation of atomic Dicke states of
Ref.\cite{Stockton2004} the control $u$ is the strength of a
magnetic field, which in an experiment cannot assume arbitrarily
large values.

Of course there is a range of useful cost and bequest functions
which cannot be cast into an affine form. For example, the variance
of a certain observable, the von Neumann entropy of the state of a
quantum system, or sub-system, the purity of a quantum system
characterized by the trace of the square of the density matrix, and
various entanglement measures are important quantities used to
characterize desirable properties of quantum systems, e.g., in
precision metrology and quantum information theory.

As described above, if we aim to produce a definite pure target state $\ket{\psi _{T}}$,
we will maximize the expectation value of the particular pure state projector
$P_{T}=\ket{\psi_{T}}\bra{\psi _{T}}$, and we thus have a linear bequest function.
If, however, we only wish to maximize the purity, but we do not care precisely which state is
produced, for a given $\varrho $, we could look for the nearest pure state, and maximize the
expectation value of the corresponding projection operator, and since that projector now depends
on $\varrho$, we effectively obtain a non-linear bequest function.

The search for "the nearest pure state" can be parametrized by the ''post-processing'' $u(T)$
dependence of a linear bequest observable, and we can more generally write the minimization of
quantum state controlled functionals of the type (\ref{J}) as
\begin{equation*}
\mathsf{S}\left[ \varrho\right] =\inf_{u}\left\{ \left\langle \varrho,G\left(
u\right) \right\rangle :u\in\mathcal{U}(T)\right\}.
\end{equation*}

\textbf{Example}: Consider the affine bequest function with $\vec{u}\in\mathcal{U}=\mathbb{R}^3$,
\begin{equation*}
\mathsf{G}\left(u,\varrho\right) =O_{B_{1}}^{+}\left( \vec{u}\right) -\vec{q%
}\cdot\vec{u},
\end{equation*}
of the qubit state $\varrho=1-\sigma_{\vec{q}}$ corresponding to the
generalized qubit cost observable $G\left( \vec{u}\right)
=O_{B_{1}}^{+}\left( \vec{u}\right)I +\sigma_{\vec{u}}$, including the
constraint function $O_{B_{1}}^{+}\left( \vec{u}\right) $ for the unit ball $%
B_{1}=\left\{ \mathbf{u}\in\mathbb{R}^{d}:\left\vert \mathbf{u}\right\vert
\leq1\right\} $ in $d\leq3$. Then $\mathsf{S}\left[ \varrho\right]
=\inf_{\vec{u}}\left\langle \varrho,G\left( \vec{u}\right) \right\rangle $ is the
closed concave function
\begin{equation}
\mathsf{S}\left[ \varrho\right] =\inf_{\vec{u}\in\mathbb{R}^{3}\,}\left\{
\,\left\langle \varrho,\sigma_{\vec{u}}\right\rangle +O_{B_{1}}^{+}\left(
\vec{u}\right) \right\} =-\sup_{\mathbf{u}\in B_{1}}\vec{q}\cdot \mathbf{u}%
=-\left\vert \mathbf{q}\right\vert ,
\label{eq:Sexp}
\end{equation}
where $\mathbf{q}$ is the projection of $\vec{q}\in\mathbb{R}^{3}$ onto $%
\mathbb{R}^{d}$. In this way we recover the concave bequest
function $\mathsf{S}\left( T,\varrho\right) =-\left\vert \mathbf{q}%
\right\vert $ used as a measure of purity by Wiseman and Bouten for $d=2$ in Ref.~\cite{Wiseman2008}.


\section{Quantum dynamics under observation}
\label{sec:HJB-obs}

\subsection{Quantum measurements and posterior states}

The state of an individual continuously measured quantum system does not
coincide with the solution of the deterministic ME (\ref{master equation}),
but instead depends on the random measurement output $y^t_\omega$ in a causal
manner. The posterior $\varrho_{\bullet}^{t}$ density matrix should be viewed
as an $\mathcal{S}$-valued stochastic process
$\varrho_{\bullet}^{t}:\omega\mapsto\varrho_{\omega}^{t}$, causally depending
on the particular observations $y_{\omega}^{t}=\left\{ y_{\omega}\left( r\right) :r<t\right\}$,
which are, in turn, obtained with a probability distribution determined by
the previous posterior states $\left\{\varrho_{\bullet}^{r}:r<t\right\}$.
Here the symbol $\bullet$ denotes a random variable, when its actual value
$\omega$ is not displayed.

The causal dependence of the posterior state $\varrho _{\bullet }^{t}$ on the
measurement data $y_{\bullet }^{t}$ is given by a corresponding quantum filtering
equation derived in the general form by Belavkin in \cite{Belavkin1988},
\cite{Belavkin1989b,Belavkin1990}. The quantum trajectories, introduced by Carmichael
\cite{Carmichael1993}, and the Monte Carlo Wave Functions (MCWF), introduced by Dalibard,
Castin and M\o lmer \cite{Dalibard1992,Molmer1993}, are stochastic pure state
descriptions of dissipative quantum systems. In these approaches the dissipative
coupling to a reservoir and resulting mixed state dynamics of a small quantum
system is ``unravelled'' by simulated Gedankenmeasurements on the reservoir.
These descriptions are included in Belavkin's formulation, which, however, does
not assume a complete detection of all reservoir degrees of freedom, and hence
it retains the density matrix description. More importantly, however, it does not
only deal with the simulation of the unavoidable dissipation of a quantum system,
but also with the dynamics induced by the probing of the system by coupling to a measurement
apparatus. One may, for example, probe atomic internal state populations and coherences
in a single atom, or a collection of atoms, by the phase shift or rotation of field
polarization experienced by a laser beam interacting with the atoms. This measurement
may be turned on and off, and several measurements may go on simultaneously as controlled
by the field strengths of different probing laser beams. Here for simplicity we display
only the diffusive case corresponding to homodyne or heterodyne detection in optics.
These detection schemes were identified as continuous limits of the Monte Carlo Wave Function
quantum jump dynamics, associated with photon counting experiments with strong local oscillator
fields \cite{Castin1992,Wiseman1993a}.

The quantum diffusive filtering equation as derived in \cite{Belavkin1988,Belavkin1989b}
for probing by coupling to a single set of system observables $L,L^\dagger$ has the form
\begin{equation}
\mathrm{d}\varrho _{\bullet }^{t}+\upsilon \left( \varrho _{\bullet
}^{t}\right) \,\mathrm{d}t=\theta\left( \varrho _{\bullet }^{t}\right)
\,\mathrm{d}w\left( t\right),
\label{filteqw}
\end{equation}
where the time coefficient $\upsilon $ contains the commutator with the Hamiltonian and the
damping terms in the deterministic ME (\ref{master equation}).
The right hand side of the equation, where $dw(t)$ denotes an infinitesimal standard Wiener Gaussian
process with $dw^2(t) = dt$, provides the fluctuation innovation term,
\begin{equation}
\mathrm{d}w\left( t\right) \equiv
\mathrm{d}y_{\bullet}\left( t\right) - \left\langle \varrho_{\bullet}^{t},L+L^{\dagger }\right\rangle \mathrm{d}t,
\label{innovation}
\end{equation}
governed by the difference between the random outcome of the measurements and its expectation value.
This term acts on the density operator as specified by
\begin{equation}
\theta\left( \varrho \right) = L\varrho +\varrho L^{\dagger
}-\left\langle \varrho ,L+L^{\dagger }\right\rangle \varrho.
\label{fluctcoefw}
\end{equation}
In optical homodyne detection the term $\mathrm{d}y_{\omega}\left( t\right)$ in Eq.(\ref{innovation}) describes
the continuous photocurrent, which is the output signal obtained from the detector.

Hereafter we shall use negative integers $j_-$ to indicate the damping dissipative operators $L^{j_-}$
and positive integers $j_+$ to describe the dissipative operators $L^{j_+}$ due to the coupling
of the system with the measurement apparatus. The same notation will be applied to the drift term
$\upsilon(\varrho) = \sum_{j_-}\upsilon_{j_-}(\varrho) + \upsilon_0(\varrho) + \sum_{j_+}\upsilon_{j_+}(\varrho)$,
where
\begin{equation}
\upsilon_0(\varrho) = \frac{\rm i}{\hbar}[H,\varrho],
\label{eq:v0}
\end{equation}
\begin{equation}
\upsilon_{j_{\pm}}(\varrho) =
\frac{(L^{j_{\pm}})^\dagger L^{j_{\pm}}\varrho+\varrho (L^{j_{\pm}})^\dagger L^{j_{\pm}}}{2}
- L^{j_{\pm}}\varrho (L^{j_{\pm}})^\dagger.
\label{eq:vjpm}
\end{equation}

\textbf{Example:} Assume an undamped qubit system with vanishing Hamiltonian,
and consider the probing described by the observable
$L^1=\frac{\lambda }{2}\sigma _{z}\equiv L$ with $\lambda\in\mathbb{R}$.
Using the Bloch vector notation for the system density matrix we can write
$L\varrho +\varrho L^{\dagger }=\lambda \left( \sigma _{z}+z\right) $ and
$\left\langle \varrho ,L+L^{\dagger }\right\rangle =\lambda z$, where
we used the fact that $\vec{q}=-\left( x,y,z\right)=-\vec{r}$. Therefore, the drift
term $\upsilon_1$ and the fluctuation coefficient $\theta$ in the filtering
equation (\ref{filteqw}) are given by:
\begin{equation*}
\upsilon_1 \left( \varrho \right) =
\frac{\lambda^2}{2}\sigma_{\vec{r}^{\perp}_{\vec{e}_z}},
\end{equation*}
\begin{align*}
\theta\left( \varrho \right) &=\lambda \left[ \left( 1-z^{2}\right)
\sigma _{z}-z\sigma_{\vec{r}^{\perp}_{\vec{e}_z}} \right] \\
\phantom{=} &= \lambda \left[ \left( 1-z^{2}\right)
\sigma _{z}-z\left(x\sigma_x + y\sigma_y\right) \right],
\end{align*}
where $\vec{r}^{\perp}_{\vec{n}} = \vec{r}-(\vec{r}\cdot \vec{n})\vec{n}$
with $\vec{n}$ being a vector in $\mathbb{R}^3$ of unit norm. In our specific
case $\vec{n}=\vec{e}_z = (0,0,1)$, and $\vec{r}^{\perp}_{\vec{e}_z} = (x,y,0)$.
The fluctuation coefficient can also be rewritten as
$\theta\left( \varrho \right) =\sigma _{\vec{l}}$, where
$\vec{l}=\lambda \left( -xz,-yz,1-z^{2}\right)$. The innovation process
driving the qubit filtering equation is defined by
$\mathrm{d}y_{\omega}\left( t\right) -\lambda \left\langle \varrho _{\omega }^{t},\sigma _{z}\right\rangle \mathrm{d}t$.

\subsection{Average change of stochastic functionals}

Let $\left\{ \varrho _{\omega }^{r}\left( t,\varrho \right) :\omega \in
\Omega \right\} $ be the solution of (\ref{filteqw})
for $r>t$\ starting in state $\varrho _{\omega }^{t}=\varrho $ at time $r=t$
for all $\omega \in\Omega$.
Then, for a smooth functional $\mathsf{F}$ on $\mathcal{L}$, we have the average
rate of change
\begin{equation*}
\lim_{h\searrow 0}\frac{1}{h}\left\{ \mathbb{E}\left[ \mathsf{F}\left[ \varrho _{\bullet }^{t+h}\left(
t,\varrho \right) \right] -\mathsf{F}\left[ \varrho \right] |\varrho \right] \right\}
=D\left( t,\varrho \right) \mathsf{F}\left[ \varrho \right],
\end{equation*}
where $\mathbb{E}\left[ \cdot|\varrho\right]$ denotes the average of
a functional of the stochastic state $\varrho$ at time
$t$. Since the change in $\varrho$ contains both deterministic
terms, linear in $dt$, and fluctuating terms, scaling with
$\sqrt{dt}$, we apply the It\^o rule \cite{Oksendal2000} and expand
the function to second order in small variations to get the correct
average rate of change. Hence, the elliptic operator $D\left(
t,\varrho \right)$, in the diffusive case is

\begin{align}
D\left( t,\varrho \right) \mathsf{F}\left[ \varrho \right] =
-\left\langle \upsilon \left( t,\varrho\right) ,\nabla _{\varrho }\mathsf{F}\left[
\varrho \right] \right\rangle +\frac{1}{2} \Delta_{\varrho }\mathsf{F}\left(t,\varrho \right),
\label{Dw}
\end{align}
where the It\^o correction is given by

\begin{align}
\Delta_{\varrho }\mathsf{F}\left(t,\varrho \right) = \left\langle \theta\left( t,\varrho \right)
^{\otimes 2},\nabla_{\varrho }^{\otimes 2}\mathsf{F}\left[ \varrho \right] \right\rangle.
\label{itocor}
\end{align}

For an $N$ level system, whose state can be described by a
generalized Bloch vector $\mathbf{r}$ in $\mathbb{R}^{N^2 - 1}$, the
notation $\nabla _{\varrho }^{\otimes 2}$ reads as $\nabla
_{\mathbf{r}}^{\otimes
2}\equiv(\nabla_{\mathbf{r}})\nabla_{\mathbf{r}}^{\sf{T}}$, where
$\nabla_{\mathbf{r}}$ is the $N^2 - 1$ column gradient vector
operator and $(\nabla_{\mathbf{r}})^{\sf{T}}$ is its transpose. The
same notation applies for the operator $\theta\left( t,\varrho
\right)^{\otimes 2}$. 

\textbf{Example:} 
Let us illustrate the above expression for a functional
$\mathsf{F}\left[ \varrho \right]=f(r)$, where $r=|\vec{r}|$ is the
length of the Bloch vector.

We consider again the situation of the previous example, where $L^1=\frac{\lambda }{2}\sigma _{z}\equiv L$.
Then, we obtain
\begin{equation*}
\left\langle \upsilon_1\left( \varrho \right), \nabla _{\varrho }\mathsf{F}\left[\varrho \right]\right\rangle =
\frac{\lambda^2}{2}\vec{r}_{\vec{e}_z}^{\perp}\cdot \vec{\nabla}f(r)
\end{equation*}
where we used the result of the third example in Sec. \ref{sec:sgd} for
$\nabla _{\varrho }\mathsf{F}\left[\varrho \right]$.
Then, the operator $\theta\left(\varrho \right)^{\otimes 2}$ can be written as the matrix
\begin{equation*}
\theta\left(\varrho \right) ^{\otimes 2} \equiv
\vec{l}\otimes\vec{l}=\lambda^2
\left(
\begin{array}{c}
-zx \\
-zy \\
1-z^{2}
\end{array}
\right)
\left(
\begin{array}{c}
-zx \\
-zy \\
1-z^{2}
\end{array}
\right)^{\intercal},
\end{equation*}
with $\vec{l}$ given in the previous example, and 
$\nabla_{\varrho }^{\otimes 2}\mathsf{F}\left[ \varrho \right]$
can be identified with the Hessian matrix $\mathfrak{H}[f](\vec{r})$
as discussed in the third example of Sec. \ref{sec:sgd}.
The It\^o correction (\ref{itocor}) is hence given by

\begin{equation*}
\frac{\Delta _{\varrho }\mathsf{F}\left( t,\varrho \right)}{\lambda^2} =\left(
\begin{array}{c}
-zx \\
-zy \\
1-z^{2}
\end{array}
\right)^{\intercal }\left(
\begin{array}{ccc}
f_{xx} & f_{xy} & f_{xz} \\
f_{yx} & f_{yy} & f_{yz} \\
f_{zx} & f_{zy} & f_{zz}
\end{array}
\right) \left(
\begin{array}{c}
-zx \\
-zy \\
1-z^{2}
\end{array}
\right),
\end{equation*}
where $f_{xy}=\frac{\partial ^{2}f}{\partial x\partial y}$.

Since $f$ depends only on $r$ and $\vec{\nabla}f(r) = r^{-1}\vec{r}\partial_r f$, the Hessian matrix
$\mathfrak{H}[f](\vec{r})$ can rewritten as

\begin{equation*}
\mathfrak{H}[f](\vec{r}) = \frac{1}{r}\frac{\partial f}{\partial r}I
+ \frac{1}{r^2}\left(\frac{\partial^2f}{\partial r^2}
-\frac{1}{r}\frac{\partial f}{\partial r}\right)
\left(
\begin{array}{ccc}
x^2 & xy & xz \\
yx & y^2 & yz \\
zx & zy & z^2
\end{array}
\right).
\end{equation*}
Hence, we have

\begin{equation*}
\Delta _{\varrho }\mathsf{F}\left( t,\varrho \right) =\frac{\vert\vec{l}\vert^2}{r}
\frac{\partial f}{\partial r}+
\frac{\lambda^2}{r^2}\left(\frac{\partial^2f}{\partial r^2}
-\frac{1}{r}\frac{\partial f}{\partial r}\right)z^2(1 - r^2)^2,
\end{equation*}
and the elliptic operator in (\ref{Dw}) becomes
\begin{align*}
D\left( t,\varrho \right) \mathsf{F}\left[ \varrho \right] & =
\frac{\lambda^2}{2r}\frac{\partial f}{\partial r}
\left(\vec{q}_{\vec{e}_z}^{\perp}\cdot\vec{r}+\frac{\vert\vec{l}\vert^2}{\lambda^2}\right)\\
\phantom{=} & +\frac{\lambda^2}{2 r^2}\left(\frac{\partial^2f}{\partial r^2}
-\frac{1}{r}\frac{\partial f}{\partial r}\right)z^2(1 - r^2)^2.
\end{align*}
For instance, if $f(r) = 1 - r^2$, then the second line in the above equation disappears and
therefore $D\left( t,\varrho \right) \mathsf{F}\left[ \varrho \right] = \lambda^2(r^2-1)(1-z^2)$.

Simultaneous probing of different observables, represented by operators $L^{j_+}$, 
coupled for example to different probing light beams, leads to a vector of random
measurement outputs, and is governed by the filtering equation

\begin{align}
\mathrm{d}\varrho _{\bullet }^{t}+\upsilon \left( \varrho _{\bullet
}^{t}\right) \,\mathrm{d}t 
= \sum_{j_+=1}^{n}\theta^{j_+}\left( \varrho
_{\bullet }^{t}\right) \,\mathrm{d}w_{j_+}(t)
\label{filteqwm}
\end{align}
and with

\begin{equation}
\theta^{j_+}\left( \varrho \right) =L^{j_+}\varrho
+\varrho (L^{j_+})^{\dagger }-\langle \varrho ,L^{j_+}
+(L^{j_+})^{\dagger }\rangle
\varrho.
\label{fluctcoefwm}
\end{equation}

Note that both the measurement induced terms and the diffuse
operation term contain the relevant probing strengths through the magnitude
of the operators $L^{j_+}$. In the following section we shall treat these
measurement strengths as our control parameters, and see how a system is
optimally controlled by measurements alone.

\textbf{Example:} In the above example we showed the explicit case of probing $\sigma_z$ with a
coupling strength $\lambda$. By cyclic permutation of the coordinates $(x,y,z)$ we obtain the
equivalent expressions for probing along the other coordinate axes, and by continuous rotation of
$(x,y,z)$ the effect of probing along an arbitrary direction can be derived.

Assuming $L^{\vec{n}}=\frac{\lambda }{2}\sigma _{\vec{n}}$ with $\vec{n}$ of unit norm, we can easily
generalise the previous results:

\begin{equation*}
\left\langle \upsilon_{\vec{n}}\left( \varrho \right),
\nabla _{\varrho }\mathsf{F}\left[\varrho \right]\right\rangle =
\frac{\lambda^2}{2r}\frac{\partial f}{\partial r}\left[r^2
- (\vec{n}\cdot\vec{r})^2\right]
\end{equation*}

$\theta(\varrho) = \sigma_{\vec{l}}$ with $\vec{l}=\lambda\left[\vec{n}-(\vec{n}\cdot\vec{r})\vec{r}\right]$,
and

\begin{equation*}
\Delta _{\varrho }\mathsf{F}\left( t,\varrho \right) =\frac{\vert\vec{l}\vert^2}{r}
\frac{\partial f}{\partial r}+
\frac{\lambda^2}{r^2}\left(\frac{\partial^2f}{\partial r^2}
-\frac{1}{r}\frac{\partial f}{\partial r}\right)(\vec{n}\cdot\vec{r})^2(1-r^2)^2.
\end{equation*}
Hence, the elliptic operator (\ref{Dw}) reads

\begin{align*}
D\left( t,\varrho \right) \mathsf{F}\left[ \varrho \right] & =
\frac{\lambda^2}{2r^2}(1-r^2)\left\{r \frac{\partial f}{\partial r}
\left[1-(\vec{n}\cdot\vec{r})^2\right]\right.\\
\phantom{=} & \left.+\left(\frac{\partial^2f}{\partial r^2}
-\frac{1}{r}\frac{\partial f}{\partial r}\right)(\vec{n}\cdot\vec{r})^2(1-r^2)\right\},
\end{align*}
where $\vert\vec{l}\vert^2$ has been written explicitly. Since
$D\left( t,\varrho \right) \mathsf{F}\left[ \varrho \right]$ provides the average change of
$\mathsf{F}\left[ \varrho \right]$ for the system while monitoring $L^{\vec{n}}$, we are now able,
for any $\varrho$, to make the optimum choice of observable $L_{\vec{n}}$ which, locally in time,
gives the largest change. We want $\mathsf{F}\left[ \varrho \right]$ to decrease as fast as possible,
and hence

\begin{align}
(\vec{n}\cdot\vec{r})^2\left[(1-r^2)\frac{\partial^2f}{\partial r^2}
- \frac{1}{r} \frac{\partial f}{\partial r}\right]
+ r \frac{\partial f}{\partial r}\leq 0
\label{eq:inequality}
\end{align}
must be satisfied. If we assume that $\partial^2_r f <0$, then the coefficient in the square brackets in
(\ref{eq:inequality}) is positive when

\begin{align*}
\left\vert \frac{\partial^2f}{\partial r^2}\right\vert < -\frac{1}{r(1-r^2)}\frac{\partial f}{\partial r},
\end{align*}
which implies that $\partial_r f$ must be negative, i.e., $f(r)$ is monotonic. Then, the minimum is reached
for $(\vec{n}\cdot\vec{r})=0$. When we consider the function $f(r) = 1 - r^2$, the above conditions are satisfied
and it tells us that it is optimal to perform a measurement in the orthogonal direction $\vec{n}$ with respect
to the state $\vec{r}$, when our objective is the purification of the state. This conclusion was also obtained
by Jacobs in Ref.\cite{Jacobs2003}.

Note that in the example we showed that a proper choice among measurements leads to the highest local
increase of the purity, but we remind that this may not necessarily be the optimal way to obtain high purity
of the final quantum state after a total probing time $T$. In the following section we shall treat the
measurement strengths as our control parameters, and apply the Bellman principle to identify how a
system is optimally controlled by measurements alone.


\section{Optimal Feedback Control with Constraints}
\label{sec:ofc}

We assume a quantum system under continuous observation described by
the filtering equation (\ref{filteqw}). A choice of the control function
$\left\{ u\left( r\right) :r\in \lbrack t_{0},t)\right\} $ is required before we
can solve the filtering equation (\ref{filteqw}) at the time $t$ for a given initial
state $\varrho _{0}$ at time $t_{0}$. To this end, we define the optimal average cost on the
interval $\left[t_0,T\right] $ to be
\begin{equation}
\mathsf{S}\left( t,\varrho\right) :=\inf_{\left\{ u_{\bullet}
\right\} }\,\mathbb{E}\left[ \mathsf{J}_{\bullet}\left[ \left\{
u_{\bullet}\left( t\right) \right\} ;t_0,\varrho\right] \right] ,  \label{S}
\end{equation}
where the minimum is considered over all admissible measurable
control strategies $\left\{ u_{\bullet}\left( t\right) :t\geq
t_0\right\}$ adapted with respect to the innovation process in
Eq.(\ref{innovation}). By admissible we mean any stochastic process
$u_{\omega}(t)$ for which the controlled filtering equation is well
defined and has a unique solution $\varrho_{\omega}^t$ (for more
precise mathematical definitions see, for example,
\cite{Oksendal2000}). The aim of feedback control theory is then to
find an optimal control strategy $\left\{
u_{\bullet}^{\mathrm{o}}\left( t\right) \right\} $ and evaluate
$\mathsf{S}\left( t,\varrho\right) $ on a fixed time interval
$[t_{0},T]$.

\subsection{Optimality equation for observed systems}
\label{sec:optEqnStoch}

We consider the problem of computing the minimum average cost in (\ref{S}).
Even though the cost is random, the Bellman principle can be
applied also in this case. As before, we let $\left\{u_{\omega }^{\mathrm{o}}\left( t\right) \right\}$
be a stochastic control leading to optimality and let
$\varrho _{\omega }^{\mathrm{o}}\left(r\right)$ be the corresponding state
trajectory (now a stochastic process) starting from $\varrho $ at time $t$.
Again choosing $t<t+h<T$, we have by the Bellman principle
\begin{gather*}
\mathbb{E}\left[ \mathsf{S}\left( t+h,\varrho _{\bullet }^{\mathrm{o}}\left(
t+h\right) \right) \right] +o\left( h\right) =\mathsf{S}\left( t,\varrho
\right) \\
+\inf_{u}\,\left\{ \frac{\partial \mathsf{S}}{\partial t}\left( t,\varrho
\right) +\mathsf{C}\left( u,\varrho \right) +D\left( u,\varrho \right)
\mathsf{S}\left( t,\varrho \right) \right\} h.
\end{gather*}
Taking the limit $h\rightarrow 0$ yields the quantum backward Bellman
equation
\begin{equation}
-\dfrac{\partial \mathsf{S}}{\partial t}\left[ \varrho \right]
=\inf_{u}\left\{ \mathsf{C}\left( u,\varrho \right) +D\left( u,\varrho
\right) \mathsf{S}\left[ \varrho \right] \right\}
\label{bellman}
\end{equation}
as derived in \cite{Belavkin1988},\cite{Belavkin1989d}.

This can be rewritten in the generalized HJB form as
\begin{equation}
\frac{\partial }{\partial t}\mathsf{S}\left( t,\varrho \left( q\right)
\right) =\mathsf{\mathsf{H}}_{\upsilon }^{\theta }\left( q,\nabla _{\varrho
}^{\otimes }\mathsf{S}\left( t,\varrho \left( q\right) \right) \right)
\label{HJBs}
\end{equation}
in terms of the generalized (Bellman) "Hamiltonian" which takes in the
diffusive case the second order derivative form
\begin{align*}
\text{\textsf{$\mathsf{H}$}}_{\upsilon }^{\theta }\left( q,\nabla _{\varrho
}^{\otimes }\mathsf{S}\right) &:=\sup_{u\in \mathcal{U}}\left\{\left\langle
\upsilon \left( u,\varrho \left( q\right) \right) ,\nabla _{\varrho }\mathsf{S}
\right\rangle -\mathsf{C}\left( u,\varrho \left( q\right) \right) \right.\\
\phantom{=}&\left.-\frac{1}{2}%
\Delta _{\varrho } \mathsf{S}\left( u,\varrho\right)\right\} ,
\end{align*}
where $\Delta _{\varrho }\mathsf{S}\left( u,\varrho \right)$ is defined in
(\ref{itocor}). This equation is to be solved backwards with the terminal condition
$\mathsf{S}\left( T,\varrho \right)=\mathsf{S}_{T}\left( \varrho \right)$.

If $\theta$ in (\ref{filteqw}) and (\ref{fluctcoefw}) does not depend on $u$
(the control is only in $H$ and not in $L^{j_-},L^{j_+}$), this gives the diffusive
HJB equation with a possible nonlinear dependence only on the first derivative
$\nabla_{\varrho } \mathsf{S}\left[ \varrho \right]$:
\begin{equation*}
-\frac{\partial \mathsf{S}}{\partial t}+\mathsf{\mathsf{H}}_{\upsilon
}\left( q,\nabla _{\varrho }\mathsf{S}\right) =\frac{1}{2}\left\langle
\theta\left( \varrho \right) ^{\otimes 2},\nabla _{\varrho }^{\otimes 2}
\mathsf{S}\right\rangle .
\end{equation*}
Exactly as in the deterministic case, the solution $\mathsf{S}$ of
this diffusive equation defines the optimal strategy through
$\nabla_{\varrho} \mathsf{S}\left( t,\varrho \right)$.

In the case we control the strength of different kinds of
measurements carried out on the system, $\theta^{j_+}$ and the
associated drift term components depend on the control parameters
$u$. Then, the optimality equation is nonlinear only in the Hessian

\begin{equation}
\frac{\partial \mathsf{S}}{\partial t}=
\langle \upsilon_0\left(\varrho \right),\nabla _{\varrho }\mathsf{S}\rangle
+\mathsf{H}^{\theta}_{\emptyset}\left( q,\nabla _{\varrho }\mathsf{S},
\nabla _{\varrho}^{\otimes 2}\mathsf{S}\right),
\label{HJBmeasw}
\end{equation}
where $\upsilon _{0}$ is defined in (\ref{eq:v0}), and $\mathsf{H}^{\theta}_{\emptyset}$
is defined later in Eq.(\ref{eq:Hthetanot}).
Here we assume the possibility that we can control also the dissipative channels,
and therefore their drift terms $\upsilon_{j_-}(\varrho)$ are included in the
Hamiltonian $\mathsf{H}^{\theta}_{\emptyset}$.


\section{Optimal control of purification}
\label{sec:oep}

\subsection{Generalized Bellman Hamiltonian}
\label{sec:ocp}

A pure Hamiltonian control does not change the eigenvalues of
quantum states and therefore does not change the entropy of any
state as a natural bequest function of the purification. On the
other hand, the filtering dynamics (\ref{filteqw}) changes the
entropy, as it provides the state conditioned on measurements. Thus,
both a Hamiltonian feedback together with the continued probing of
the system, and a feedback strategy where new measurements are
selectively carried out on the system, can be applied to optimize
the convergence towards a pure state. The former possibility has
been studied by Jacobs \cite{Jacobs2003}, Wiseman and Bouten
\cite{Wiseman2008}. Instead, we shall assume control of the coupling
$\lambda$ to the continuous measurement in the filtering equation as
a real function of time $t$ and the information previously obtained.
Thus, we are interested in the optimal purification strategy via the
feedback measurement control in one or several channels by solving
the optimality equation (\ref{HJBmeasw}) with the Hamiltonian
$\mathsf{H}^{\theta}_{\emptyset}$ containing explicitly all the
diffusive measurement terms (\ref{fluctcoefwm}) and the
corresponding dissipation drifts $\upsilon _{L^{j_+}}$. We allow,
however, also the control of the dissipative operators which
correspond to some unobserved modes with velocity term
$\upsilon_{j_-}$.

We may assume that the operators $L^{j}=\lambda ^{j}R^{j}$, with $j$ being either $j_+$ or
$j_-$, are controlled only by the strengths $u_{j}=\left\vert \lambda ^{j}\right\vert^{2}$
or by the phases $u_{j}=\arg \lambda^{j}$ of the coupling parameters
$\lambda^{j}$ with fixed measurement operators $R^{j}$.
Taking the first (controlled strength) choice with $\arg \lambda ^{j}=0$, we
define the corresponding $\upsilon _{j}=u_{j}\upsilon _{R^{j}}$,
$\theta^{j}=\sqrt{u_{j}}\theta _{R^{j}}$ and
$\Delta _{\varrho }^{j}\mathsf{S}=\left\langle \theta _{R^{j}}^{\otimes 2},
\nabla _{\varrho }^{\otimes 2}\mathsf{S}\right\rangle $ in terms of the rescaled
$\upsilon _{R^{j}}$ and $\theta _{R^{j}}$. We then obtain
\begin{equation}
\label{eq:Hthetanot}
\mathsf{H}^{\theta }_{\emptyset}:=\sup_{\{u_{j}\leq 0\}}\left\{ \sum_{j\neq 0}u_{j}
\left( \left\langle \upsilon_{R^{j}},\nabla _{\varrho }\mathsf{S}\right\rangle
-\frac{1}{2}\Delta_{\varrho }^{j}\mathsf{S}\right) -\mathsf{C}\left( \mathbf{u}\right)
\right\} ,
\end{equation}
explicitly in terms of the measurement strengths $u_{j}$.

Thus, we have a convex optimization problem under the constraint $u^{j}\geq 0$. We will consider the
optimization problem under the further natural constraint of a given maximum total probing strength,
$\left\Vert \mathbf{u}\right\Vert _{1}:=\sum_{j}u^{j}\leq 1$, which is incorporated by choosing the
cost function $\mathsf{C}\left( \mathbf{u},\varrho \right) =+\infty $ if $\left\Vert
\mathbf{u}\right\Vert _{1}>1$, and $\mathsf{C}\left( \mathbf{u},\varrho \right) =0$
otherwise. Equation (\ref{eq:Hthetanot}) reduces to
\begin{equation}
\mathsf{H}^{\theta}_{\emptyset}=\max_{j}\left\{ \left\langle \upsilon
_{j}\left( \varrho \right) ,\nabla _{\varrho }\mathsf{S}\left( t,\varrho
\right) \right\rangle -\frac{1}{2}\Delta _{\varrho }^{j}\mathsf{S}\left(
t,\varrho \right) \right\}
\label{optmeasHam}
\end{equation}
if at least one value under the maximum is positive, and $u_{_{j}}^{\mathrm{o}}\left( t\right) =1$
for any optimal $j=j_{\mathrm{o}}\left(\varrho ,\mathsf{S}\right)$ and
$u_{_{j}}^{\mathrm{o}}\left( t\right) =0$ $\forall j\neq j_{\mathrm{o}}$. Otherwise, $\mathsf{
\mathsf{H}}_{\emptyset }^{\theta }=0$, no measurement
purifies $\varrho$, and the maximum is achieved on the
optimal feedback strategy $\mathbf{u}^{\mathrm{o}}\left( t\right) =0$.
In the case of a single
channel measurement this defines a simple two-valued strategy for when the
probing should be switched on and off, $u\in \{0,1\}$, corresponding to the
Hamiltonian
\begin{equation}
\mathsf{\mathsf{H}}_{\emptyset }^{\theta }=\left\vert \left\langle \upsilon
_{1}\left( \varrho \right) ,\nabla _{\varrho }\mathsf{S}\left( t,\varrho
\right) \right\rangle -\frac{1}{2}\Delta _{\varrho }^1\mathsf{S}\left(
t,\varrho \right) \right\vert _{+},
\label{dfHam}
\end{equation}
where $\left\vert x\right\vert _{+}=\max \left\{ 0,x\right\}$.

\subsection{Purifying a qubit only with measurements}
\label{sec:vb}

Let us take the cost $c\left( \mathbf{u}\right)
=O_{\mathcal{U}}^{+}\left( \mathbf{u}\right)$ of the constraint
$\mathcal{U}=\left\{ u^{j}\geq 0:\sum_{j}u^{j}\leq 1\right\}$ and
the hermitian operators
$L^{\vec{n}_\Omega}=\frac{\lambda(\Omega)}{2}\sigma_{\vec{n}_{\Omega}}$,
where instead of the integer $j_+$ we use the outward normal unit
vector, $\vec{n}_{\Omega}$, parametrized by the continuous solid
angle argument, $\Omega$, along which the diffusive measurement is performed.
In the following we set $\lambda(\Omega)=\sqrt{u(\Omega)} = 1$, as
discussed in the previous section. Besides, we assume that the
system is subject to no dissipation and no Hamiltonian control. In
this case, the diffusive filtering equation (\ref{filteqw}) reduces
to

\begin{align}
\mathrm{d}\varrho _{\bullet }^{t}+\upsilon_{\Omega} \left( \varrho _{\bullet }^{t}\right)
\,\mathrm{d}t = \theta^{\Omega}\left( \varrho _{\bullet}^{t}\right)
\,\mathrm{d}w\left( t\right).
\label{eq:SMEqubitPur}
\end{align}
This equation can be rewritten in terms of the state vector $\vec{r}$ as

\begin{align*}
\left(
\begin{array}{c}
\mathrm{d}x\\
\mathrm{d}y\\
\mathrm{d}z
\end{array}
\right)
+
\left(
\begin{array}{c}
x-(\vec{n}\cdot\vec{r})n_x\\
y-(\vec{n}\cdot\vec{r})n_y\\
z-(\vec{n}\cdot\vec{r})n_z
\end{array}
\right)\frac{\mathrm{d}t}{2} =
\left(
\begin{array}{c}
n_x-(\vec{n}\cdot\vec{r})x\\
n_y-(\vec{n}\cdot\vec{r})y\\
n_z-(\vec{n}\cdot\vec{r})z
\end{array}
\right)\mathrm{d}w,
\end{align*}
where we used the results of the examples in Sec. \ref{sec:HJB-obs}. 
As noticed in the last example of Sec. \ref{sec:HJB-obs}, about the purification of
a qubit state, the elliptic operator (\ref{Dw}) attains its largest negative value if
$\vec{n}\cdot\vec{r} = 0$, that is we should observe the system in a Bloch vector direction
orthogonal to the one of the state density matrix. This simplifies the diffusive filtering
equation considerably.

In our example, the HJB equation 
for the optimal control of qubit purification (in its very compact
form) is given by:

\begin{widetext}
\begin{align*}
&-\frac{\partial \mathsf{S}}{\partial t}
+\frac{1}{2}\int_{4\pi}\mathrm{d}\Omega\,\delta(\vec{n}_{\Omega}-\vec{n}_{\Omega_{\mathrm{o}}})
\left(\vec{r}_{\vec{n}_{\Omega}}^{\perp }\cdot \vec{\nabla}\mathsf{S}
\right) \\
& =\frac{1}{2}\int_{4\pi}\mathrm{d}\Omega\,\delta(\vec{n}_{\Omega}-\vec{n}_{\Omega_{\mathrm{o}}})\left\{
\alpha_{\Omega}^2\left[
\vec{\nabla}\left(\vec{r}^{\perp}_{\vec{n}_{\Omega}}\cdot\vec{\nabla}\mathsf{S}\right)
\cdot\vec{r}^{\perp}_{\vec{n}_{\Omega}}
-\vec{r}_{\vec{n}_{\Omega}}^{\perp }\cdot\vec{\nabla}\mathsf{S}
\right]
+\left( 1-\alpha_{\Omega}^{2}\right) \left[
\left( 1-\alpha_{\Omega}^{2}\right) \frac{\partial
^{2}\mathsf{S}}{\partial \alpha_{\Omega}^{2}}
-2 \alpha_{\Omega} (\vec{r}_{\vec{n}_{\Omega}}^{\perp }\cdot \vec{\nabla}\mathsf{S}_{\vec{n}_{\Omega}})
\right]
\right\},
\end{align*}
\end{widetext}
where $\delta(\vec{n}_{\Omega}-\vec{n}_{\Omega_{\mathrm{o}}})$ is the Dirac delta function on
the surface of the unit sphere, $\alpha_{\Omega} = \vec{n}_{\Omega}\cdot\vec{r}$, and
$\mathsf{S}_{\vec{n}_{\Omega}}=\frac{\partial\mathsf{S}}{\partial \alpha_{\Omega}}$.
If $\alpha_{\Omega} = 0$ is chosen, then $\vec{r}^{\perp}_{\vec{n}_{\Omega}} =
\vec{r}-\alpha_{\Omega}\vec{n}_{\Omega}=\vec{r}$ and the above equation simplifies to
\begin{align*}
-\frac{\partial \mathsf{S}}{\partial t}
+\frac{1}{2}\vec{r}\cdot \vec{\nabla}\mathsf{S} = 0.
\end{align*}
If we write $\vec{r} = r(\sin\varphi\sin\vartheta,\cos\varphi\sin\vartheta,\cos\vartheta)$ this equation
can be reduced to
\begin{align}
-\frac{\partial \mathsf{S}}{\partial t}
+\frac{r}{2}\frac{\partial\mathsf{S}}{\partial r} = 0.
\label{eq:optEqPur}
\end{align}
It is straightforward to check that $\mathsf{S}(t,r) = 1 - r^2 e^{-(T-t)}$ solves (\ref{eq:optEqPur})
with $\mathsf{S}(T,r) = 1 - r^2$. Note that this solution is obtained under the assumption that we
always measure a Bloch-sphere component orthogonal to the current density matrix Bloch vector.
It is, however, easy to verify that the supremum according to (\ref{optmeasHam}) is in accord with
that choice. This confirms the demonstration by Wiseman and Bouten in Ref.\cite{Wiseman2008} of the
optimality of Jacobs \cite{Jacobs2003} purification protocol.

\subsection{Controlling only the fluctuations $\theta^{j_+}$}

As a special constraint on our control consider $\mathbf{u}=\left(u_{j}\right)$ indexed
by $j_{\pm }=\pm j$ and add the constraints $u_{j_{-}}=1-u_{j_{+}}$ with $R^{j_{-}}=R^{j_{+}}$
for all $j=1,\ldots ,n$ under the constraint $\sum_{j\geq 1}u_{j}\leq 1$ (and $\sum_{j}u_{j}=n$).
The new constraint corresponds to keeping the dissipation drifts for each pair $\left( -j,j\right)$
independent of the controls, $\upsilon _{-j}+\upsilon _{+j}=\upsilon _{R^{j}}$ and the optimality
equation reduces to
\begin{equation}
\frac{\partial \mathsf{S}}{\partial t}=\left\langle \upsilon \left( \varrho
\right) ,\nabla _{\varrho }\mathsf{S}\right\rangle -\frac{1}{2}
\min_{j>0}\Delta _{\varrho }^j\mathsf{S}\left( t,\varrho \right).
\label{OptEq}
\end{equation}
Hence, $\mathsf{H}_{\emptyset }^{\theta }=\left\langle \upsilon\left( \varrho\right),
\nabla _{\varrho}\mathsf{S}\right\rangle +\mathsf{\mathsf{H}}$, where the new
Hamiltonian, $\mathsf{H}$, is defined by the minimal Hessian
$\Delta _{\varrho }^j\mathsf{S}\left(t,\varrho \right)$.
Equation (\ref{OptEq}) becomes linear if one of the Hessians, say 
$\Delta _{\varrho }^1\mathsf{S}\left( t,\varrho \right)$,
is the most negative, $\Delta_{\varrho }^1\mathsf{S}\leq \Delta _{\varrho }^j\mathsf{S}$
for all $j$, $\varrho $ and $t$.

The constraint $u_{j_{-}}=1-u_{j_{+}}$ with $R^{j_{-}}=R^{j_{+}}$
corresponds, for example, to the partial monitoring of a dissipative
channel which leaks information to the enviroment. Such a leakage can be, 
for example, the one of resonance
fluorescence from an atomic quantum system monitored with finite
detection efficiency or within a finite solid angle, proportional to
$\upsilon _{+j}$.

The qubit purification protocol we have discussed in Sec. \ref{sec:vb}
does not contain dissipation, as in Refs.\cite{Jacobs2003,Wiseman2008}. 
Those results would be changed in the presence of pure damping terms ($j<0$) 
and it would be more difficult to obtain a pure state. Here the choice of 
constraint $u_{j_{-}}=1-u_{j_{+}}$ allows us to reduce the diffusive filtering 
equation of the observed system to a stochastic master equation where no 
dissipative terms appear. More precisely, the equation would have only the
drift terms $\upsilon _{R^{j}}$. For instance, the dynamics of a damped qubit, with 
only a dissipative term $\upsilon_{-1}$ and a measurement observables $L^1$, 
as before, would be governed by a filtering equation formally identical
to (\ref{eq:SMEqubitPur}), but now with the possibility to control only the 
fluctuation operator $\theta^{j_+}$. Again, however, we can decide to perform 
a measurement or not by looking at the minimal Hessian $\Delta _{\varrho
}^j\mathsf{S}\left(t,\varrho \right)$. 


\section{Discussion}
\label{sec:disc}

We have presented a general formalism for the optimal control of a quantum system subject to
measurements, where the control can both be of a suitable feedback Hamiltonian and of the choice of
future measurements carried out on the system. The use of measurements to prepare and protect pure and
entangled quantum states can thus be made subject of systematic investigation, and optimal schemes can
be deviced for given physical setups.

It is the philosophy of our work that the quantum state of a controlled system, i.e., its von Neumann
density matrix, is treated in the same way as classical control engineers treat the state of their
classical systems. The Bellman principle can then be applied in in the same way as for classical
states. In the present work we derived the corresponding Hamilton-Jacobi-Bellman theory for a wider
class of controls and cost functionals than traditionally considered in the literature.

Another interesting problem, which is explicitly solvable but is formulated in the infinite
dimensional Hilbert space, is the setup problem for the quantum feedback control with ``soft''
constraints given by quadratic cost functions $c$ and $g$ in $u$ \cite{Belavkin1987,Belavkin1988},
\cite{Belavkin1983}, \cite{Belavkin1979}. It reduces to the linear optimal control problem in the
finite-dimensional space for the sufficient coordinates of the Gaussian Bosonic states exactly as
in the classical linear-quadratic Gaussian case. For a more detailed discussion with proofs we refer
to \cite{Belavkin1999}, and for a particular case in a more recent work by Yanagisawa \cite{Yanagisawa2006}.


\section*{Acknowledgment}

V.P.B. would like to thank for the hospitality at the University of Aarhus and
support from the Lundbeck Foundation. K.M. and A.N. acknowledge financial
support from the European Union Integrated Project SCALA, and K.M.
acknowledges support of the ONR MURI on quantum metrology with atomic
systems.


\bibliography{Negretti_QuantumControl}

\end{document}